# Towards an Efficient Anomaly-Based Intrusion Detection for Software-Defined Networks.

Majd Latah [1*], Levent Toker [2]

[1] Department of Computer Science, Ozyegin University, 34794, Cekmekoy, Istanbul, Turkey
[2] Department of Computer Engineering, Ege University, 35100, Bornova, Izmir, Turkey
[*] majd.latah@ozu.edu.tr; levent.toker@ege.edu.tr

**Abstract:** Software-defined networking (SDN) is a new paradigm that allows developing more flexible network applications. SDN controller, which represents a centralized controlling point, is responsible for running various network applications as well as maintaining different network services and functionalities. Choosing an efficient intrusion detection system helps in reducing the overhead of the running controller and creates a more secure network. In this study, we investigate the performance of the well-known anomaly-based intrusion detection approaches in terms of accuracy, false alarm rate, precision, recall, f1-measure, area under ROC curve, execution time and Mc Nemar's test. Precisely, we focus on supervised machine-learning approaches where we use the following classifiers: Decision Trees (DT), Extreme Learning Machine (ELM), Naive Bayes (NB), Linear Discriminant Analysis (LDA), Neural Networks (NN), Support Vector Machines (SVM), Random Forest (RT), K Nearest-Neighbour (KNN), AdaBoost, RUSBoost, LogitBoost and BaggingTrees where we employ the well-known NSL-KDD benchmark dataset to compare the performance of each one of these classifiers.

## 1. Introduction

Network security is one of the most important aspects in modern communications. Recently, programmable networks have gained popularity due to their abstracted view of the network which, in turn, provides a better understanding of the complex network operations and increases the effectiveness of the actions that should be taken in the case of any potential threat. Software Defined Networking (SDN) represents an emerging centralized network architecture, in which the forwarding elements are being managed by a central unit, called an SDN controller, which has the ability to obtain traffic statistics from each forwarding element in order to take the appropriate action required for preventing any malicious behavior or abusing of the network. At the same time, the SDN controller uses a programmable network protocol, which is OpenFlow (OF) protocol, in order to communicate and forward its decisions to OF-enabled switches [1].

In spite of the significant impact of using a centralized controller, the controller itself creates a single point of failure, which makes the network more vulnerable compared with the conventional network architecture [2]. On the other hand, the existence of a communication between the OF-enabled switches and the controller opens the door for various attacks such Denial of Service (DoS) [3], Host Location Hijacking and Man in the Middle (MIM) attacks [4]. Therefore, in order to develop an efficient Intrusion Detection System (IDS) for SDNs, the system should be able to make intelligent and real time decisions. Commonly, an IDS designed for SDNs works on the top of the controller, which forms an additional burden on the controller itself. Thus, designing a lightweight IDS is considered advantageous, since it helps in effectively detecting of any potential attacks as well as performing other fundamental network operations such as routing and load balancing in a more flexible manner. Scalability, is also an important factor, which should be taken into consideration during the designing stage of the system [4]. There are two main groups of intrusion detection systems: signature-based IDS and anomaly-based IDS. Signature-based IDS searches for defined patterns within the analyzed network traffic. On the other hand, an anomaly-based IDS can estimate and predict the behavior of system. A signature-based IDS shows a good performance only for specified well-known attacks. On the contrary, anomaly-based IDS enjoys ability to detect unseen intrusion events, which is an important advantage for detecting zero day attacks [5].

Anomaly-based IDS can be grouped into three main categories [5]: statistical-based approaches, knowledge-based approaches, and machine learning-based approaches. In this study, we focus on machine learning-based approaches. Machine learning techniques can be categorized into four categories: (i) supervised techniques, (ii) semi-supervised techniques, (iii) unsupervised techniques and (iv) reinforcement techniques. In this paper, we investigate various supervised learning techniques with respect to their accuracy, false alarm rate, precision, recall, f1-measure, area under ROC curve, Mc Nemar's test and time taken to train and test each classifier.

## 2. Related work

Previous research efforts for providing a detailed analysis of supervised machine learning techniques used for intrusion detection are summarized in Table 1. These studies focused on training and testing different machine learning approaches using standard intrusion detection datasets. However, obtaining all these features from an SDN controller could be computationally expensive. Therefore, we have two possible choices: either using a subset of these standard datasets [6] or extracting new features based on network traces of standard datasets or statistics provided by the controller [7]. In this study, we use a subset of features extracted from NSL-KDD dataset based on employing the well-known Principal Components Analysis (PCA) approach and considering the following supervised machine learning approaches: Decision Trees (DT), Extreme Learning Machine (ELM), Naive Bayes (NB), Linear Discriminant Analysis (LDA), Neural Networks (NN), Support Vector Machines (SVM), Random Forest (RT), Nearest-Neighbor (KNN), AdaBoost, RUSBoost, LogitBoost and BaggingTrees. As mentioned before, for performance measurement, we use accuracy, false alarm rate, precision, recall, f1-measure and area under ROC curve, as well as time taken to train and test each one of these classifiers. Furthermore, we use Mc Nemar's test in order to statistically demonstrate that a significant increase has been achieved by an algorithm over the other one.

**Table 1** Overview of previous supervised machine learning studies for intrusion detection

| Ref. | Year | Algorithms | Dataset |
|---|---|---|---|
| [8] | 2005 | C 4.5 | KDD CUP'99 |
| | | K-nearest neighbor | |
| | | Multi-layer perceptron | |
| | | Regularized discriminant analysis | |
| | | Fisher linear discriminant | |
| | | Support vector machines | |
| [9] | 2007 | Decision Trees | KDD CUP'99 |
| | | Random Forest | |
| | | Naive Bayes | |
| | | Gaussian classifier | |
| [10] | 2009 | J48 | NSL-KDD |
| | | Naive Bayes (NB) | |
| | | NB Tree | |
| | | Random Forest | |
| | | Random Tree | |
| | | Multi-layer perceptron (MLP) | |
| | | SVM | |
| [11] | 2010 | Discriminative multinomial Naïve Bayes classifiers | NSL-KDD |
| [12] | 2013 | Principal component analysis based feature selection, Genetic algorithm based detector generation, J48, NB, MLP, BF-Tree, NB- Tree, RF Tree. | NSL-KDD |
| [13] | 2013 | Correlation based feature selection and consistency based filtering, ADTree, C4.5, J48graft, LADTree, NBTree, RandomTree, RandomForest, REPTree | NSL-KDD |
| [14] | 2013 | J48, BayesNet, Logistic, SGD, IBK, JRip, PART, Random Forest, Random Tree and REPTree | NSL-KDD |
| [15] | 2015 | Neural Networks | NSL-KDD |
| [16] | 2016 | Logistic Regression Gaussian Naive Bayes SVM and Random Forest | NSL-KDD |

## 3. Dataset

As mentioned earlier, in this study we use NSL-KDD dataset. NSL-KDD is an improved version of KDD Cup99 dataset, which suffers from huge number of redundant records [10]. Both KDD Cup99 and NSL-KDD datasets include the features shown in Table 2. It is worth mentioning that these features fall into four different categories as described in Table 3.

**Table 2** List of features of KDD Cup '99 dataset.

| F. # | Feature name. | F. # | Feature name. | F. # | Feature name. |
|---|---|---|---|---|---|
| F1 | Duration | F15 | Su attempted | F29 | Same srv rate |
| F2 | Protocol type | F16 | Num root | F30 | Diff srv rate |
| F3 | Service | F17 | Num file creations | F31 | Srv diff host rate |
| F4 | Flag | F18 | Num shells | F32 | Dst host count |
| F5 | Source bytes | F19 | Num access files | F33 | Dst host srv count |
| F6 | Destination bytes | F20 | Num outbound cmds | F34 | Dst host same srv rate |
| F7 | Land | F21 | Is host login | F35 | Dst host diff srv rate |
| F8 | Wrong fragment | F22 | Is guest login | F36 | Dst host same src port rate |
| F9 | Urgent | F23 | Count | F37 | Dst host srv diff host rate |
| F10 | Hot | F24 | Srv count | F38 | Dst host serror rate |
| F11 | Number failed logins | F25 | Serror rate | F39 | Dst host srv serror rate |
| F12 | Logged in | F26 | Srv serror rate | F40 | Dst host rerror rate |
| F13 | Num compromised | F27 | Rerror rate | F41 | Dst host srv rerror rate |
| F14 | Root shell | F28 | Srv rerror rate | F42 | Class label |

**Table 3** List of feature categories presented in NSL-KDD dataset

| Category | Features |
|---|---|
| Basic features | F1,F2,F3,F4,F5,F6,F7,F8,F9,F10 |
| Content features | F11,F12,F13,F14,F15,F16,F17,F18 F19,F20,F21,F22 |
| Time-based features | F23,F24,F25,F26,F27,F28,F29,F30,F31 |
| Host-based features | F32,F33,F34,F35,F36,F37,F38,F39,F40 F41 |

As shown in Table 4, NSL-KDD includes a total of 39 attacks where each one of them is classified into one of the following four categories (DoS, R2L, U2R, Probe). Moreover, a set of these attacks is introduced only in the testing set. These new attacks are indicated in bold font.

**Table 4** List of attacks presented in NSL-KDD dataset

| Attack category | Attack name |
|---|---|
| Denial of service (DoS) | **Apache2**, Smurf, Neptune, Back, Teardrop, Pod, Land, **Mailbomb**, **Processtable**, **UDPstorm** |
| Remote to local (R2L) | WarezClient, Guess_Password, WarezMaster, Imap, Ftp_Write, **Named**, MultiHop, Phf, Spy, **Sendmail**, **SnmpGetAttack**, **SnmpGuess**, **Worm**, **Xsnoop**, **Xlock** |
| User to root (U2R) | Buffer_Overflow, **Httptuneel**, Rootkit, LoadModule, Perl, **Xterm**, **Ps**, **SQLattack** |
| Probe | Satan, **Saint**, Ipsweep, Portsweep, Nmap, **Mscan** |

In addition, Table 5 shows the distribution of the normal and attack records in NSL-KDD training and testing sets.

**Table 5** Distributions of attacks and normal records in NSL-KDD dataset

| | Total Records | Normal | DoS | R2L | U2R | Probe | Total Records |
|---|---|---|---|---|---|---|---|
| KDD Train | 125973 | 67343 53.46% | 45927 36.46 | 995 0.79% | 52 0.04% | 11656 9.25% | |
| KDD Test | 22544 | 9711 43.07% | 7458 33.08% | 2754 12.22% | 200 0.89% | 2421 10.74% | |

## 4. Feature Selection

In order to increase the efficiency of SDN based intrusion detection systems we need to select the best features that can be used in SDN context. It is worth noting that the content features need to be omitted due to the fact that these features are complex to extract by a network based IDS [17]. Therefore, content features (i.e. F11 to F22) were excluded from the NSL-KDD dataset. For the remaining features we apply Principal Component Analysis (PCA) on the training set. PCA allow us to transform a large dataset into a new, smaller and uncorrelated one [18]. The standard approach of PCA can be summarized in the following 6 steps [19]:

- Find the covariance matrix of the normalized $d$-dimensional dataset.
- Find the eigenvectors and eigenvalues of the covariance matrix.
- Sort the eigenvalues in descending order.
- Select the $k$ eigenvectors that correspond to the $k$ largest eigenvalues.
- Construct the projection matrix from the $k$ selected eigenvectors.
- Transform the original dataset to obtain a new $k$-dimensional feature space.

In this study, we employ PCA in the following steps:
- First, we extract the features with the largest coefficients from the principal components.
- Second, we select the $k$ eigenvectors that correspond to the $k$ largest eigenvalues.
- Third, we transform the original dataset with corresponding features using the projection matrix from the k selected eigenvectors.
- Finally, we validate the performance of the selected features and corresponding $k$ component by applying the decision tree approach on the training test.

## 5. Evaluation Metrics

The performance of each classifier is evaluated in terms of accuracy, False Alarm Rate (FAR), precision, recall, F1-measure, Area Under ROC Curve (AUC), execution time and Mc Nemar's test. A good IDS should achieve high level of accuracy, precision, recall and F1-measure with low false alarm rate. The accuracy is calculated by:

$$Accuracy = \frac{TP+TN}{(TP+TN+FN+FP)} \quad (1)$$

True Positives (TP) is the number of attack records correctly classified; True Negatives (TN) is the number of normal traffic records correctly classified; False Positives (FP) is the number of normal traffic records falsely classified and False Negatives (FN) is number of attack records instances falsely classified. False alarm rate is calculated by:

$$False\ Alarm\ Rate = \frac{FP}{TN+FP} \quad (2)$$

We also calculate the precision, recall and F1-measure for each classifier where precision is calculated by:

$$\Pr ecision = \frac{TP}{TP+FP} \quad (3)$$

Recall is also calculated by:

$$\operatorname{Re} call = \frac{TP}{TP+FN} \quad (4)$$

And F1-measure is calculated by:

$$F1-measure = 2 \times \frac{(\Pr ecision \times \operatorname{Re} call)}{(\Pr ecision + \operatorname{Re} call)} \quad (5)$$

In addition, we evaluate the performance of the previously selected classifiers based on their execution time as well as the analysis of the receiver operator characteristic (ROC) curve, where the area under curve (AUC) can be used to compare each classifier with another one. The higher AUC, the better IDS. One other important metric that can be used for comparing two algorithms is Mc Nemar's test, which is a non-parametric pair-wise test shows that a statistically significant increase has been achieved by an algorithm over the other one. When z-value of Nemar's Test > 1.96 ($p$-value is less than 0.05), the conclusion is that there is a significant difference between the two algorithms. Z-score is used to show the confidence levels [22].

$$z = \frac{(|N_{12} - N_{21}|) - 1}{\sqrt{(N_{12} + N_{21})}} \quad (6)$$

$N_{12}$: represents the number of times when the first algorithm success in classification and other one fails.
$N_{21}$: represents the number of times when the second algorithm success in classification and the first one fails.

## 6. Experimental Results

The experiment is conducted on Intel i5 machine with 12 GB of RAM. As shown in Fig 1, we get the best results when selecting 9 of the top features that contribute to the all PCA's components as input, which need to be transformed to less dimensional space of the corresponding components.

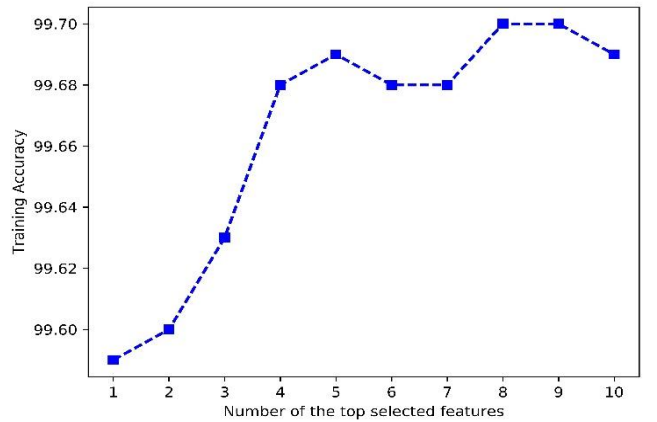

**Fig. 1.** The level of accuracy obtained by using top selected features

These 9 selected features are: F27, F30, F5, F23, F8, F1, F2, F39, F3. Brief description of these features is provided in Table 6.

**Table 6** List of feature selected from NSL-KDD dataset

| Feature | Description |
|---|---|
| F27 | Percentage of connections that have REJ errors |
| F30 | Percentage of connections to different services |
| F5 | Number of data bytes from source to destination |
| F23 | Number of connections to the same host as the current connection in the past two seconds |
| F8 | Number of wrong fragments |
| F1 | Duration of the connection in seconds |
| F2 | Connection protocol (tcp, udp, icmp) |
| F39 | Percentage of connections to the current host and serror rate specified service that have an S0 error |
| F3 | Destination port mapped to service |

Fig. 2, shows the level of accuracy achieved when using different number of principal components. The best results achieved with the first 10 components.

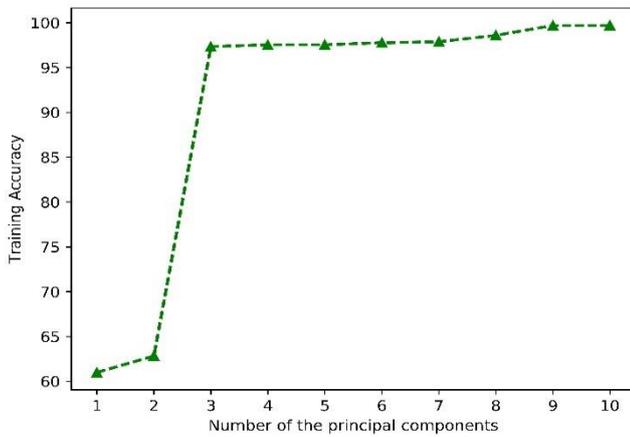

**Fig. 2.** The level of accuracy obtained with different number of PCA's principle components

Table 7 shows the results obtained for both training and testing stages. In terms of accuracy level, the most accurate classifiers for the training stage are: DT, RF, BaggingTrees, RUSBoost and AdaBoost with a slight difference between them. For the testing stage, however, we notice that DT approach achieved the highest level of accuracy followed by AdaBoost, RUSBoost and BaggingTrees. One can observe that ensemble methods achieved a lower false positive rate compared to DT.

**Table 7** Detection accuracy and false alarm rate obtained after training and testing different supervised machine learning algorithms with 10 principle components

| Method | Accuracy (%) | | False Alarm Rate (%) | |
|---|---|---|---|---|
| | Training | Testing | Training | Testing |
| Naive Bayes | 64.16 | 49.12 | 4.54 | 5.74 |
| LDA | 72.37 | 70.32 | 9.98 | 3.76 |
| Linear SVM | 91.04 | 81.40 | 9.21 | 5.92 |
| NN | 92.15 | 74.23 | 4.54 | 6.38 |
| ELM | 92.66 | 75.86 | 5.54 | 3.57 |
| KNN | 98.14 | 82.31 | 1.92 | 3.53 |
| LogitBoost | 98.95 | 84.85 | 0.94 | **2.83** |
| AdaBoost | 99.03 | 87.16 | 1.03 | 3.68 |
| RUSBoost | 99.19 | 85.57 | 0.96 | 3.59 |
| BaggingTrees | 99.33 | 84.03 | 0.81 | 3.51 |
| RandomForest | **99.70** | 80.13 | **0.29** | 3.49 |
| Decision Tree | **99.70** | **88.74** | 0.31 | 3.99 |

In terms of false alarm rate, it is worth mentioning that LogitBoost approach achieved the best results. Therefore, one can conclude that ensemble methods such AdaBoost and LogitBoost can achieve a good accuracy with low false positive rate.

From both Table 7 and Table 8, one can observe that using PCA feature selection enhanced the accuracy level for most of the classifiers in compared with using the basic features provided by the SDN controller (F1, F2, F5, F6, F23 and F24). In terms of area under ROC curve, as shown in Fig. 3 (a), we notice that DT and RF approaches achieved the best AUC for the training task followed by BaggingTrees, RUSBoost, AdaBoost and LogitBoost with slight difference between each other. Both NN and SVM had nearly the same AUC for the training task. NB, however, achieved the least training AUC.

**Table 8** Detection accuracy and false alarm rate obtained after training and testing different supervised machine learning algorithms based on basic features provided by the SDN controller (i.e. features number F1, F2, F5, F6, F23 and F24).

| Method | Accuracy (%) | | False Positive Rate (%) | |
|---|---|---|---|---|
| | Training | Testing | Training | Testing |
| Naive Bayes | 59.27 | 49.88 | 3.7227 | 5.14 |
| LDA | 87.57 | 69.36 | 3.26 | 2.24 |
| SVM | 90.86 | 71.00 | 6.55 | 10.27 |
| NN | 84.10 | 66.22 | 2.41 | 1.61 |
| ELM | 93.16 | 74.17 | 2.25 | 2.31 |
| KNN | 98.23 | 77.09 | 3.128 | 4.07 |
| RandomForest | 98.09 | 75.96 | **0** | **0** |
| Decision Trees | 98.37 | 74.43 | 0.306 | 6.43 |
| LogitBoost | 99.38 | 79.44 | 0.43 | 2.75 |
| BaggingTrees | 99.54 | 79.16 | 0.47 | 3.26 |
| AdaBoost | 99.56 | 78.94 | 0.384 | 2.76 |
| RUSBoost | **99.68** | **80.31** | 0.29 | 3.48 |

For the testing task, as shown in Fig. 3(b), one can observe that the best AUC obtained by DT followed by AdaBoost, RUSBoost, LogitBoost and BaggingTrees. KNN achieved a better AUC than SVM and RF. In the same context, we notice that SVM also achieved a higher AUC than ELM approach. In terms of precision and F1-measure the best results were achieved by DT, whereas LogitBoost achieved the best results in terms of recall.

**Table 9** Precision, Recall, F1-measure obtained after training and testing different supervised machine learning algorithms with 10 principle components

| Method | Precision (%) | Recall (%) | F1-measure (%) |
|---|---|---|---|
| Naive Bayes | 14.95 | 77.49 | 25.06 |
| LDA | 50.7 | 94.69 | 66.07 |
| Linear SVM | 71.81 | 94.13 | 81.47 |
| NN | 59.56 | 92.5 | 72.46 |
| ELM | 60.29 | 85.71 | 73.98 |
| KNN | 71.59 | 96.41 | 82.17 |
| LogitBoost | 75.53 | **97.24** | 85.03 |
| AdaBoost | 80.23 | 96.65 | 87.67 |
| RUSBoost | 77.41 | 96.6 | 85.95 |
| BaggingTrees | 74.61 | 96.56 | 84.17 |
| RandomForest | 67.73 | 96.25 | 80.13 |
| Decision Tree | **83.24** | 96.50 | **89.38** |

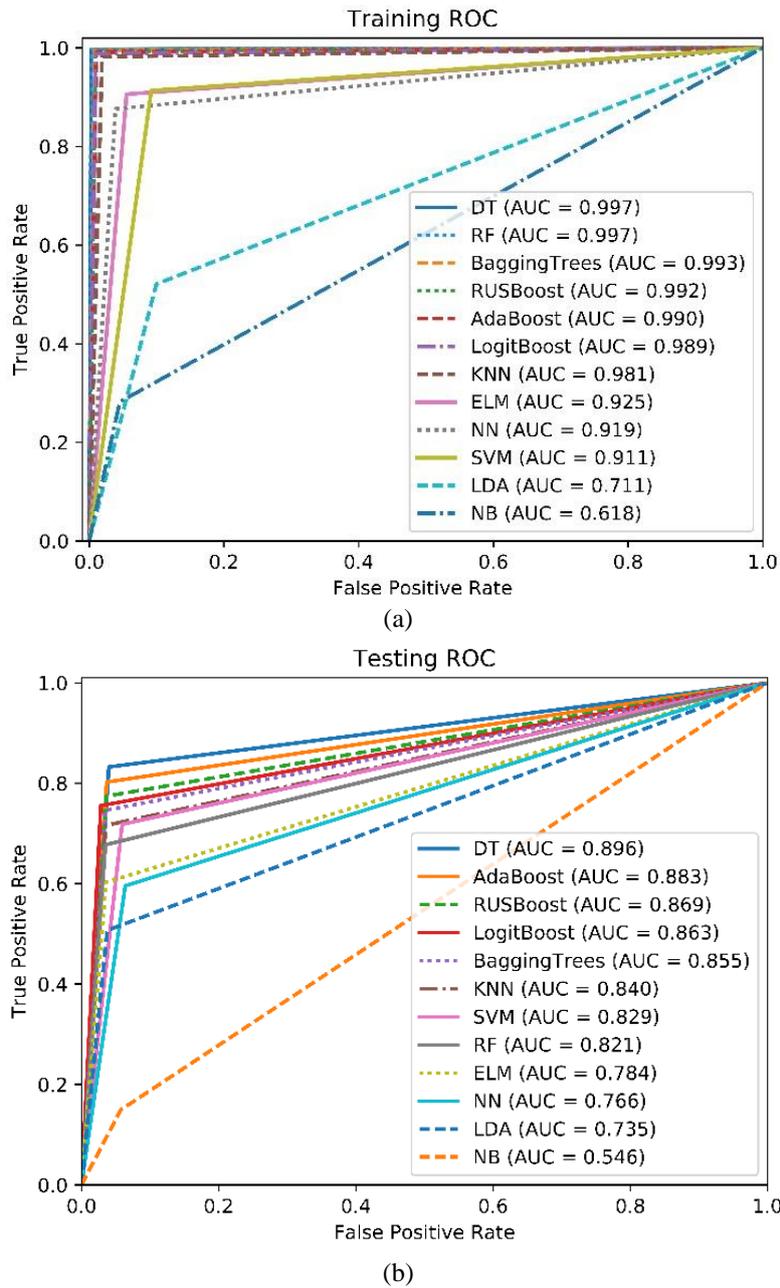

**Fig. 3.** ROC curve comparison for (a) training and (b) testing different supervised machine learning based IDS

**Table 10** Z-score values of Mc-Nemar's Test for the supervised machine-learning algorithms used in the this study (the arrowheads ← ↑ denote which classifier performed better). The shaded ones that larger than 1.96 indicate statistically significant differences at the confidence level of 95% (p < 0.05).

|        | NB      | DT      | AdaB    | RUSB    | LogitB  | Bagging | RF      | KNN     | ELM     | NN      | SVM     | LDA |
|--------|---------|---------|---------|---------|---------|---------|---------|---------|---------|---------|---------|-----|
| NB     | -       |         |         |         |         |         |         |         |         |         |         |     |
| DT     | 86.2 ←  | -       |         |         |         |         |         |         |         |         |         |     |
| AdaB   | 88.2 ←  | 7.1 ↑   | -       |         |         |         |         |         |         |         |         |     |
| RUSB   | 82.4 ←  | 19.2 ↑  | 10.3 ↑  | -       |         |         |         |         |         |         |         |     |
| LogitB | 86.2 ←  | 16.5 ↑  | 18.2 ↑  | 4.3 ↑   | -       |         |         |         |         |         |         |     |
| Bagging| 84.4 ←  | 20.7 ↑  | 22.1 ↑  | 9.2 ↑   | 6.4 ↑   | -       |         |         |         |         |         |     |
| RF     | 75.0 ←  | 40.0 ↑  | 35.7 ↑  | 30.0 ↑  | 25.2 ↑  | 24.8 ↑  | -       |         |         |         |         |     |
| KNN    | 78.1 ←  | 28.0 ↑  | 23.8 ↑  | 17.5 ↑  | 13.9 ↑  | 10.1 ↑  | 12.3 ←  | -       |         |         |         |     |
| ELM    | 71.1 ←  | 42.6 ↑  | 43.2 ↑  | 38.6 ↑  | 35.6 ↑  | 33.5 ↑  | 13.1 ↑  | 29.3 ↑  | -       |         |         |     |
| NN     | 64.4 ←  | 49.4. ↑ | 47.6 ↑  | 43.9 ↑  | 39.0 ↑  | 37.9 ↑  | 24.1 ↑  | 30.2 ↑  | 13.7 ↑  | -       |         |     |
| SVM    | 75.5 ←  | 25.1 ↑  | 22.9 ↑  | 15.7 ↑  | 14.0 ↑  | 10.5 ↑  | 4.6 ↑   | 3.4 ↑   | 17.4 ↑  | 22.3 ↑  | -       |     |
| LDA    | 58.5 ←  | 56..7 ↑ | 54.9 ↑  | 49.7 ↑  | 50.7 ↑  | 47.1 ↑  | 33.8 ↑  | 41.1 ↑  | 26.6 ↑  | 12.4 ↑  | 36.0 ↑  | -   |

For Mc Nemar's test, the null hypothesis suggest that different classifiers perform similarly whereas the alternative hypothesis claims that at least one of the classifiers performs differently. As shown in Table 10, by looking at the z-score values of Mc Nemar's Test, one can conclude that DT achieved significantly better results than the other classifiers where the alternative hypothesis was accepted with a confidence level more than 99.5%. KNN also performed better than RF. AdaBoost also performed better than the other algorithms except DT, with a confidence level more than 99.5%. Bagging and boosting produced better results over other conventional machine learning methods such as KNN, ELM, NN, RF, SVM and LDA.

In terms of execution time, as shown in Fig. 4(a), we notice that NB approach achieves the best results for the training task. We excluded KNN from Fig. 4(a) due to the fact that KNN has no training time, where this algorithm employs a distance function in order to predict the corresponding labels [20]. From Fig. 4(b), on the other hand, one can observe that ELM approach achieved the best testing time. Moreover, ELM has achieved an acceptable false alarm rate as shown in Table 7. Therefore, ELM and its improved hierarchical approach [21] can possibly be an efficient choice for SDNs.

On the other hand, in spite of the good level of accuracy for the testing stage achieved by KNN approach, it showed the worst testing time, which may indicate that KNN algorithm is not the best choice for SDNs where each controller may need to handle thousands of flows per second. A possible solution to this problem can be achieved by reducing the number of the training instance by applying an appropriate sampling method. Finally, one can observe that DT has achieved the highest level of accuracy and a good testing time in compared with the other classifiers.

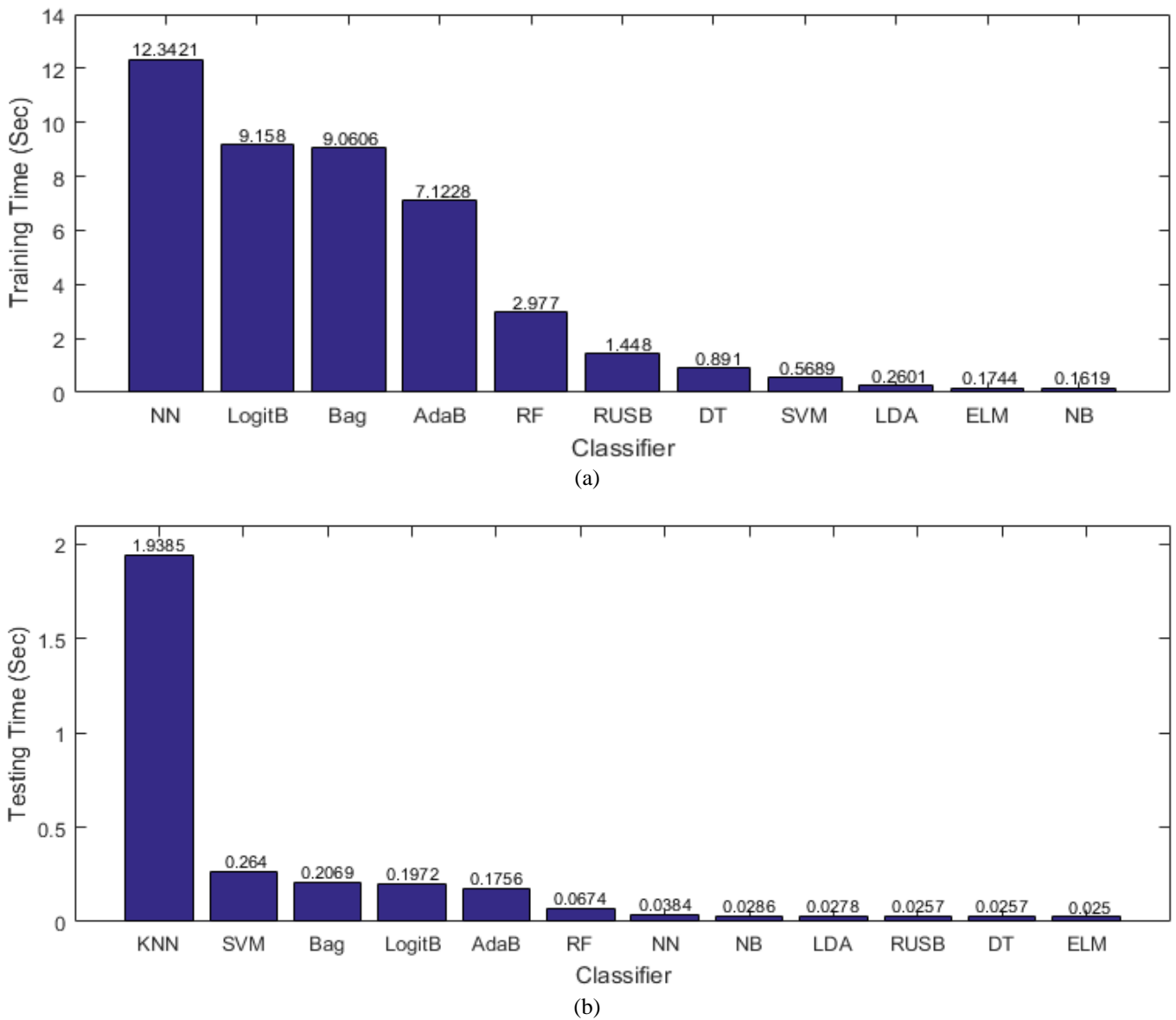

**Fig. 4.** Execution time for (a) training and (b) testing different supervised machine learning method

## 7. Conclusion

In this paper, we provide a comparative study of choosing an efficient anomaly-based intrusion detection method for SDNs. We focused on supervised machine learning approaches by using the following classifiers: NN, LDA, DT, RF, Linear SVM, KNN, NB, ELM, AdaBoost, RUSBoost, LogitBoost and BaggingTrees. In addition, we used PCA method for feature selection and dimensionality reduction. Using NSL-KDD dataset and based on our extensive experimental study, we conclude that DT approach shows the best performance in terms of accuracy, precision, F1-measure, AUC and Mc Nemar's Test. Also bagging and boosting approaches outperformed other conventional machine learning methods such as KNN, ELM, NN, RF, SVM and LDA with a confidence level more than 99.5%. Whereas in terms of false alarm rate and recall the best results achieved by LogitBoost. In terms of the execution time, ELM approach achieved the best testing time.

It is worth noting that using PCA approach was very successful in enhancing the accuracy level from 77.09% to 88.74% in compared with using the basic features provided by the SDN controller. Our future work will be focused on comparing the results obtained from this study with other machine learning approaches and exploring other flow-based features that could be used to achieve a higher level of accuracy with lower false alarm rate.

## 8. Acknowledgement

We would like to thank the reviewers for their insightful comments to improve the quality of this paper.